# Novel Ternary Ag$^{II}$Co$^{III}$F$_5$ Fluoride: Synthesis, Structure and Magnetic Characteristics

Daniel Jezierski[a]*, Zoran Mazej[b]*, Wojciech Grochala[a]*

We present a new compound in the silver-cobalt-fluoride system, featuring paramagnetic silver (d$^9$) and high-spin cobalt (d$^6$), synthesized by solid-state method in an autoclave under F$_2$ overpressure. Based on powder X-ray diffraction, we determined that Ag$^{II}$Co$^{III}$F$_5$ crystallizes in a monoclinic system with space group $C2/c$. The calculated fundamental band-gap falls in the visible range of the electromagnetic spectrum, and the compound has the character of charge-transfer insulator. AgCoF$_5$ is a ferrimagnet with one predominant superexchange magnetic interaction constant between mixed spin cations (Ag-Co) of –62 meV (SCAN result). Magnetometric measurements conducted on a powdered sample allowed the identification of a transition at 128 K, which could indicate magnetic ordering.

## Introduction

The world of transition metal fluorides is already quite rich, but a large number of even simple bimetallic systems are still awaiting to be discovered. The same applies to mixed-valence fluoride systems. For example, only three crystal structures of mixed-valence transition metal homodimetallic pentafluorides are currently known: Cr$_2$F$_5$[1], Mn$_2$F$_5$[2] and Ag$_2$F$_5$[3]. The first two, Cr$_2$F$_5$ and Mn$_2$F$_5$, crystallize in the monoclinic space group $C2/c$ and exhibit two paramagnetic metal sites, that show antiferromagnetic ordering between the spins below 40 K[1] and 53 K[2], respectively. The third, Ag$^{II}$Ag$^{III}$F$_5$, crystallizes in a triclinic structure ($P\bar{1}$) and possess diamagnetic low-spin Ag$^{III}$. In addition, one compound, Cu$_2$F$_5$[4] has been predicted theoretically, while four others, Ni$_2$F$_5$[5], Co$_2$F$_5$[6], Fe$_2$F$_5$[7] and V$_2$F$_5$[8], have been postulated as intermediates in the thermal decomposition of the corresponding TMF$_3$ (for the first three) or its hydrate (for the last one).

The family of heterobimetallic A$^{II}$B$^{III}$F$_5$ compounds, although limited in number, demonstrates remarkable diversity[1,9–16]. These includes pentafluorides with two transition metals, such as: CrTiF$_5$[1], CrVF$_5$[1], MnCrF$_5$[17], CdMnF$_5$[9]. All known examples of pentafluorides with two paramagnetic sites are limited to early transition metals, late transition metal pentafluorides have not been reported.

To date, no pentafluorides with a paramagnetic transition metal (TM) ion in the trivalent state (TM$^{III}$) have been documented within the Ag$^{II}$TMF$_5$ series. Only AgAuF$_5$, which exhibits Au in its low-spin diamagnetic trivalent state (Au$^{III}$), has been postulated[18] to be isostructural with the triclinic CuAuF$_5$ homolog. Further investigation of Ag-TM-F phases identified <u>four</u> compounds that may exhibit magnetic interactions between divalent silver (Ag$^{II}$) and transition metals (TM). These compounds include AgMn$^{IV}$F$_6$[19], which is characterized by a Curie-Weiss constant of -66 K, and others such as AgRh$^{IV}$F$_6$[20], AgRu$^V$F$_7$[21] and AgIr$^V$F$_7$[21]. However, the exact crystal structures of the Mn and Rh analogs and a detailed elucidation of the magnetic behavior of all compounds are still unresolved. This knowledge gap has initiated theoretical investigations of the magnetic coupling mechanisms between Ag and paramagnetic transition metal sin possible fluoride systems[22–24]. Remarkably, significant inter-metal magnetic superexchange constants of -45.9 meV for Ag-Cu$^{II}$ and -33.3 meV for Ag-Ni$^{II}$[22] systems have been calculated in theoretical AgTMF$_4$ models (with TM= Cu or Ni). The question naturally arises whether other TM/Ag$^{II}$ systems with even stronger magnetic interactions can be prepared.

In this study, we present a new member of the mixed-valence transition dimetal fluorides - AgCoF$_5$. The precursors of this compound – AgF$_2$ and CoF$_3$, exhibit antiferromagnetic behavior below 163 K[25] and 460 K[26], respectively. AgF$_2$ is a 2D antiferromagnet[27], while CoF$_3$ has a G-type antiferromagnetic structure[26,28]. Of these two binary fluorides, AgF$_2$ is recognized as a HTSC precursor[27,29,30], which has recently led to extensive investigations into its possible doping[31–38].

In this paper, the crystal structure, magnetic, and electronic properties of AgCoF$_5$ are described and supported by theoretical analysis.

## Results and discussion

### Synthesis

The successful synthesis of AgCoF$_5$ was achieved by a solid-state high-temperature reaction with a stoichiometric mixture of AgF$_2$ and CoF$_3$ compounds at F$_2$ overpressure, as presented in equation 1. First, the mixture was carefully ground on the mortar in a glovebox under argon atmosphere, then transferred to a nickel boat and hermetically sealed in a nickel autoclave. The autoclave was then removed from the glovebox and connected to a vacuum line. Argon was pumped away from the reaction vessel, and the system was pressurized with F$_2$ gas (up to 5 bar). The mixture underwent heating at 520°C for 24 hours. Excess fluorine prevents the thermal decomposition of Ag$^{II}$ to Ag$^I$ fluoride, and that of Co$^{III}$ to Co$^{II}$[50]. After cooling the reactor and evacuating the fluorine gas, the reactor was returned to the glovebox. The middle part of the sample was extracted from the boat to avoid contamination by nickel. The mixture was further ground and subjected to the similar conditions as before (under F$_2$ overpressure at 480 °C for 24 hours). The resulting product, a brown powder (sample designated **S1**), was characterized using various analytical methods.

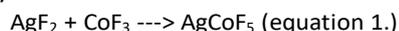

AgF$_2$ + CoF$_3$ ---> AgCoF$_5$ (equation 1.)

In our next experimental approach, we investigated the potential for the formation of a structure with mutual miscibility at the positions of metal atoms, similar to the observed behavior in AgF$_2$/CuF$_2$ mixtures[24]. To investigate this, we started the synthesis with a mixture of AgF$_2$ and CoF$_2$ at a molar ratio of 0.7:0.3 under the same conditions as in the first approach (equation 1). Upon XRD analysis, we identified the presence of two distinct phases in the final product: AgCoF$_5$ and AgF$_2$ (sample **S2**), with the respective near equimolar ratio (0.44**:**0.56) determined by Rietveld refinement (see Supporting

Information). Crucially, our measurements revealed no change in the Néel temperature ($T_N$) for $AgF_2$, which remained at 163 K. Given that the volume of the phases identified as $AgCoF_5$ and $AgF_2$ in the second approach practically coincides with the volume assigned in the first experiment (eq. 1, and also refer to Supplementary Information), we assume that solubility of the cations or partial mutual substitution at the corresponding cation sites[24] is unlikely under these conditions. This assertion is also supported by the large differences in atomic radii between $Co^{III}$ (R = 88.5 pm) and $Ag^{II}$ (R = 108 pm).

**Crystal structure of $AgCoF_5$**

The crystal structure of $AgCoF_5$ was elucidated by powder X-ray diffraction of a polycrystalline sample obtained according to equation 1 (Fig. 1). The preliminary phase analysis of the diffractogram showed the predominant presence of an unknown crystalline phase and trace amounts of $AgF_2$. All reflections corresponding to the unidentified phase in the diffraction pattern were selected for the indexing procedure using X-Cell implemented in the Materials Studio software[51]. Consequently, the logarithm suggested a monoclinic structure with a unit cell volume twice as small as shown in Table 1. By reviewing the ICSD database, we found that $Cr^{II}Cr^{III}F_5$ ($C2/c$, No. 15 Z=4) has similar diffraction pattern features to our new unknown phase. Therefore, we took the structure of $Cr_2F_5$ as a starting point for the geometry optimization of $AgCoF_5$ (using DFT+U method), substituting $Cr^{II}$ and $Cr^{III}$ atoms with $Ag^{II}$ and $Co^{III}$, respectively. Following the geometry optimization of such a structure, a Rietveld refinement was performed using the results of the theoretical calculations of $AgCoF_5$ and experimental XRD pattern (Fig. 1), resulting in good fitting parameters (Table 1).

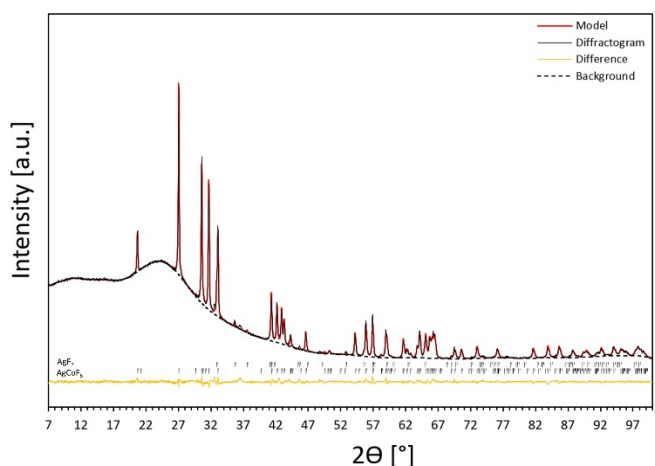

**Figure 1**. Powder XRD pattern for **S1** (black line): mainly $AgCoF_5$ (96%) with traces of $AgF_2$ (4%), next to the refined model (red) and the fitted background (as black dashed line). The black markers indicate reflex positions for the $AgF_2$ and $AgCoF_5$ model phases for the Co $K\alpha_1$ line. The differential profile is marked in orange.

Considering the final structure of $AgCoF_5$, we also performed comparative analysis between space group $C2/c$ (Z = 4) and its counterpart $C2/m$ (Z = 2). The presence of a binary rotational symmetry axis ($C2$) is observed in both groups. However, the distinct structural variations emerge from the different orientation of the symmetry plane in relation to this axis. In particular, in the monoclinic $C2/c$ space group, the symmetry plane is oriented perpendicular to the binary axis. In the $C2/m$ space group, on the other hand, the arrangement is such that the symmetry plane is aligned parallel to the two-fold rotational axis. The result of this orientation of the symmetry plane is the presence of Co-F-Co chains without any tilting (bond angle of 180 degrees), with larger interatomic distances of the metal sites along the $c$ axis compared to the $C2/c$ type. This symmetry is quite uncommon, as none of the $A^{II}B^{III}F_5$ compounds exhibit it (see *Supplementary Information*). Nevertheless, a comparative analysis of both structural forms was performed to clarify the final structure of $AgCoF_5$. The structure of $AgCoF_5$ in both proposed structural solutions is detailed in the *Supplementary Information*, where a comprehensive analysis of its features is also provided.

We have also performed a comparative computational analysis of the ground state (GS) energies and dynamic stability for both structural solutions. Our calculations demonstrate that the $C2/c$ space group is energetically more favorable, with an energy lower by 4.51 meV/FU compared to $C2/m$ (DFT+U).

Moreover, the latter displays dynamic instability, evidenced by three imaginary phonon branches, while the former has no imaginary phonons (see *Supplementary Information* for detailed information).

Indeed, if the $C2/m$ space group is assumed for the Rietveld refinement procedure using the experimental diffraction pattern, it proves to be insufficient for the description the $AgCoF_5$ structure. First, it leads to absence of several reflections – in particular at 45.7, 52.1, 52.7, 58.2 of 2θ, which are observable in the experimental data (but present for the $C2/c$ space group), and 2) the fitting parameters are much worse to those obtained for the C2/c model (see Table SI5 and Fig. SI3 in the Supplementary Information). This led us to conclude that the $C2/c$ space group accurately represents the crystal structure of the $AgCoF_5$ compound.

The key structural information of $AgCoF_5$ ($C2/c$) from the Rietveld refinement is presented in the Table. 1. Further details are provided in Table SI2.

The experimentally determined parameters of the $AgCoF_5$ unit cell were compared with the theoretical values from calculations, and the data are shown in Table SI 3. The best agreement between the Rietveld method and the theoretical values is observed for the DFT+U ($U_{Ag}$ = 5 eV) and HSE06 methods, where calculations slightly underestimate and overestimate $AgCoF_5$ volume by 0.07% and 0.34 %, respectively. For DFT+U (U = 8 eV), the volume of the unit cell is underestimated more – by 1.36%. However, the largest discrepancy is observed with the SCAN method – the volume is by 1.87% larger than experimental data.

The intermetallic distances between the paramagnetic centers are defined by the lattice vectors. Specifically, the interatomic Ag-Co distance along the a-axis is determined to be 3.814(10) Å, while the corresponding distance along the b-axis is 3.637(10) Å. The nearest Co-Co and Ag-Ag contacts oriented along [100], appear as 3.765(10) Å (more detailed and theoretical data are explained in Table SI 4). The fluorine atoms are the completing elements of the structure. They act as important bridging elements in the structure, linking metal ions, thus creating a distinct network of metal-fluorine bonds and playing a crucial role in the magnetic properties of the compound.

A salient structural motif in the $C2/c$ $AgCoF_5$ framework comprises alternating Co-F-Co chains and Ag-F-Ag rectangles aligned along the c-axis – [001] direction (Fig. 2) in the (010) plane. The bond angles in these Co-F-Co chains are less than 180° and the tilt is determined to be 162.0(3)°. In contrast, there are no chain-like Ag-F-Ag connections. Along the c crystallographic axis, the Ag-F bonds are generally the longest, as shown in Fig.4 B. The Ag---Ag connections are bridged by two fluorine atoms along [001] and form a rectangle with an Ag-F-

Ag angle of 107.6(5)°. These fluorine atoms simultaneously form covalent bonds with cobalt atoms, creating a composite structural network. Orthogonal to the Co-F-Co chains is a rectangular lattice of paramagnetic centers – silver and cobalt (001), which are bridged by fluorine atoms. The Ag-F-Co bonds have a corrugated arrangement; their angles are 158.9(12)° along the [010] direction and 127.9(8)° along the [100] direction (Fig.2 and Fig.3). All angles determined using theoretical and experimental methods are summarized in Tab. SI4 in the *Supplementary Information.*

**Table 1.** Structural parameters of AgCoF$_5$. For further information see Table ESI in the Supplementary Information.

| Parameter | AgCoF$_5$ |
|---|---|
| Crystal system | Monoclinic |
| Space group (number) | C2/c (15) |
| a, b, c [Å] | 7.274414 (2), 7.627744 (2), 7.529471 (2) |
| α=γ, β [°] | 90, 115.976(4) |
| Z | 4 |
| V [Å$^3$] | 375.580 (19) |
| Temperature, radiation type, range | 298K, Co Kα, 7° – 100° of 2Θ |
| Fit parameters | GOF = 1.66%, Rp = 1.12%, wRp = 1.68% |

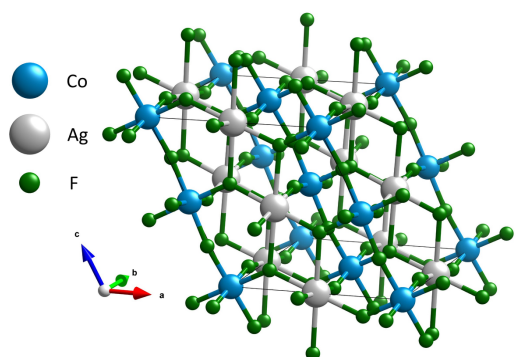

**Figure 2.** Structure of AgCoF$_5$. Color code: grey for silver, blue for cobalt and green for fluorine atoms.

The first coordination sphere of cobalt forms a distorted octahedral configuration (CN = 6), as presented in Figure 4A. Such as distortion is also observed in CoF$_3$ and can be attributed to the Jahn-Teller effect, which applies to the high-spin d$^6$ electronic configuration[28]. This distortion in AgCoF$_5$ is manifested by the presence of three distinct sets of Co-F bond lengths: 2x1.827(16) Å, 2x1.905(6) Å, and 2x1.921(13) Å. The elongation of these bonds is aligned along specific crystallographic axes, with the longest Co-F bond along the [100] direction, the second longest along [001], and the shortest along the [010] b-axis. A comparative analysis of the computational data, derived from density functional theory augmented with Hubbard U (DFT+U), and the Rietveld refinement results reveals a general agreement with the observed trend (ESI, Table SI 4).

In the primary coordination environment, the silver atoms exhibit a configuration that deviates strongly from the ideal octahedral symmetry (CN = 6), as can be seen in Figure 4B. This distortion, which is similar to that found in pure AgF$_2$, is primarily attributed to the pronounced Jahn-Teller effect, which is a characteristic feature of the d$^9$ electron configuration of divalent silver[25]. The manifestation of this effect is evident in the three distinct categories of Ag-F bond lengths within the coordination sphere: two bonds with a length of 2.052(16)Å [010], two others of 2.090(16)Å [100] and the last pair with 2.562(16)Å [001]. The computational methods provide closely matched results with a maximum deviation of 1.77% between experimental and theoretical Ag-F bond lengths in the (001) plane (HSE06 functional; for all methods see Table SI4).

The structural properties of AgCoF$_5$ described above have a profound influence on the vibrational spectra of this compound (phonons) as well as its magnetic and electronic properties. Let us take a look at the phonons.

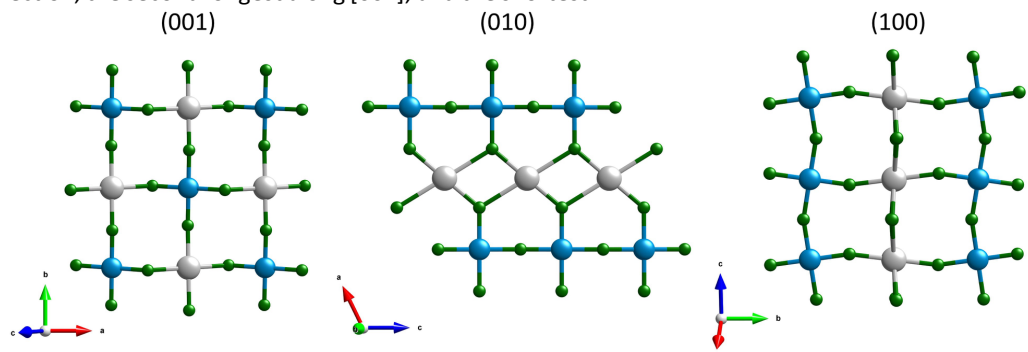

**Figure 3.** Projections of the AgCoF$_5$ structure, from left: along *c* axis (001), centre – along the *b* axis (010) and right – along the *a* axis (100) . Colour code: grey for silver, blue for cobalt and green for fluorine atoms. For further details see Table SI4 in the *Supplementary Information*.

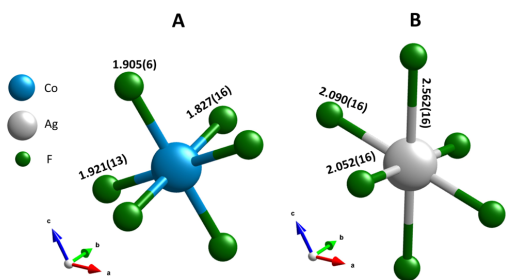

**Figure 4.** The first coordination sphere for **A.** cobalt and **B.** silver, both with CN = 6. The M-F bonds lengths are expressed in Å, following Rietveld refinement.

**Vibrational spectra of AgCoF$_5$**

In the case of the space group *C*2/*c*, the crystallographic unit cell is twice as large as the Bravais cell[52,53]. According to group theory, one should therefore expect 42 phonon vibrations of the lattice for AgCoF$_5$ (Z=2 for the primitive cell). Of these, 12 are silent (A$_u$) and 3 are translational (2B$_u$ and A$_u$). Consequently, 12 modes are infrared active: 12B$_u$, while 15 are active in Raman spectroscopy: 8B$_g$ + 7A$_g$.

Figure 5 shows comparison of the experimental IR and RAMAN spectra for AgCoF$_5$ (lines) with the theoretical positions of the bands (dashed). The assignment of the observed band positions in the spectra shown (Figure 5), along with their theoretical positions and symmetry (based on DFT+U calculations), is presented in Table SI6 in the *Supplementary Information*.

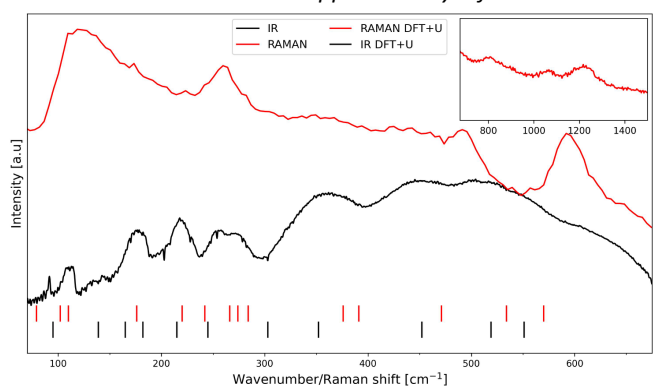

**Figure 5.** The Raman spectrum (red line) and the infrared spectrum (black line) together with the positions of the phonon vibrations based on DFT+U calculations (red and black dashes) (for exact positions see Table SI6.). An extended range for Raman spectroscopy measurement is shown in the upper right corner of the image.

In general, we find a very good agreement between the vibrational positions for the AgCoF$_5$ lattice as predicted by computational methods and the positions in the experimental spectra. This is evidenced by a high correlation coefficient $R^2$ = 0.9976 (Figure 6). In addition, the wavenumbers of the translational (acoustic) modes are calculated with an error of +/- 2 cm$^{-1}$, suggesting that the positions of other vibrations may have similarly small deviations (Table SI6). Overall, we succeeded to assign 12 out of 15 Raman-active modes (with the exception of one A$_g$ vibration calculated at 534 cm$^{-1}$ and two low-frequency modes that fall below our bottom experimental range). Also, we assigned 9 out of 12 IR-active modes (with two absent B$_u$ vibrations calculated at 303 cm$^{-1}$ and 165 cm$^{-1}$, and one falling below our experimental range). Their absence in the spectra could be due to the low intensity of the corresponding bands or/and too much background noise.

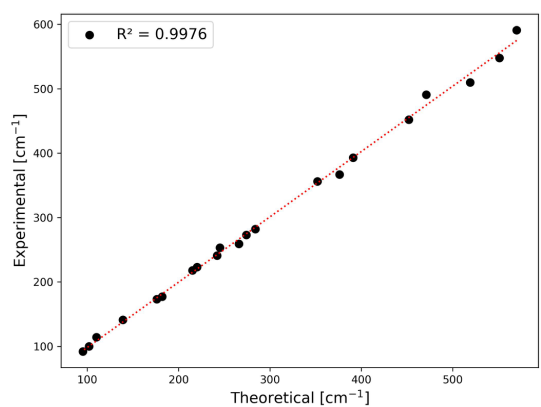

**Figure 6.** Correlation between experimental and theoretical positions for bands observed in the spectra.

**Magnetic properties of AgCoF$_5$**

To evaluate the magnetic nature of the compound, we performed magnetic measurements on a polycrystalline AgCoF$_5$ sample (**S1**) using SQIUD (Superconducting Quantum Interference Device). In addition, we performed Density Functional Theory (DFT) calculations to deepen our understanding of the magnetic behavior. Unfortunately, it was not possible to obtain samples of AgCoF$_5$ that were free of traces of AgF$_2$. The sample **S1** taken for the magnetic measurements contained about 4 mol % of AgF$_2$, as revealed by Rietveld refinement. Its presence is also confirmed by a transition in the magnetic susceptibility curve at 163 K, indicative of AgF$_2$ (in particular in the -dχ/dT curve, see Fig. 5). Thus, while we can detect the transition temperature typical of AgCoF$_5$, we decided not to determine the effective magnetic moment and the paramagnetic Curie-Weiss temperature (θ) which are strongly affected by the AgF$_2$ impurity.

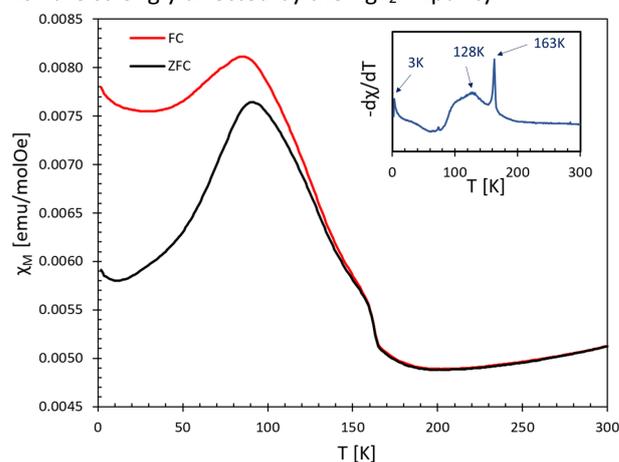

**Figure 7.** Magnetic susceptibility χ of the sample at 10 kOe. For FC regime -dχ/dT vs T shown in the inset.

Figure 7 displays the magnetic susceptibility χ(T) of the sample, measured from 2 to 300 K under both zero-field-cooled (ZFC) and field-cooled (FC) conditions at 10 kOe. The magnetic behavior of the sample is rather complex, however, based on the maximum value of -dχ/dT (FC) we determined three transition temperatures: two associated with AgCoF$_5$, $T_1$ = 128 K and $T_3$ = 3-9 K (depending on the applied field, see ESI) and one for AgF$_2$ with $T_N$ = 163 K. Between the temperatures of 300 K and 163 K, a decrease in magnetization with temperature is observed. This could indicate the existence of short-range antiferromagnetic (AFM) ordering in AgCoF$_5$ in this temperature range, similar to that observed in KAgF$_3$[54]. On the other hand, the ordering at $T_3$ is unspecific, as it may originate from a minute amount of any ferromagnetic impurities.

Below the transition temperature (128 K), a noticeable divergence in the temperature-dependent behavior of the magnetic susceptibility between the zero-field-cooled (ZFC) and field-cooled (FC) regimes can be observed. A similar feature is seen in pure $AgF_2$ samples (see Fig. SI6), where the weak ferromagnetism below the Néel temperature ($T_N$) is related to the spin-canting of silver (II), due to the Dzyaloshinskii-Moriya interaction[55,56]. A similar explanation for the weak ferromagnetism of $AgCoF_5$ below 128K can be postulated. It should be noted that the canting of $Ag^{2+}$ is likely to be more pronounced than that of $Co^{3+}$, as the former exhibits a stronger spin-orbit coupling. In the structure of $AgCoF_5$, the magnetic interactions within the square spin lattice parallel to the (001) plane are limited to the silver and cobalt atoms. In contrast, the interactions along the [001] direction are predominantly limited to homoatomic contacts, especially between two silver sites, and are similar to interlayer contacts observed in $AgF_2$.

In addition, the presence of $AgF_2$ traces with $AgCoF_5$ in **S1** can be used for comparison purposes. Thus, a significant difference in the magnetic transition widths between $AgF_2$ and $AgCoF_5$ is evident (see inset in Fig. 5). The significantly wider transition in $AgCoF_5$ in contrast to $AgF_2$, could be due to the existence of additional magnetic transitions in $AgCoF_5$ and/or to differences in the dimensionality of the magnetic ordering for both compounds. Silver(II) difluoride is identified as a two-dimensional (2D) antiferromagnet, exhibiting an exchange constant close to -70meV in the $[AgF_4]$ layers[27,33]; this leads to a sharp transition at 168K, evidenced by a distinct peak in the -$d\chi/dT$ versus T plot (see inset in Fig. 5). In the following, we focus on the M(H) dependence of the sample (**S1**).

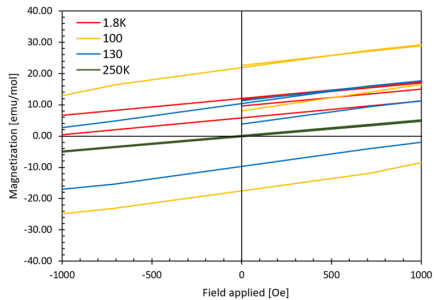

**Fig. 8** Magnetization curves at 1.8 K, 100 K, 130 K and 250 K for **S1** as a function of the applied field for fields up to 1000 Oe. The curves for the extended temperature range and the applied magnetic fields are listed in the *Supplementary Information*.

Fig. 6 shows the M(H) curve for sample **S1**. We observe hysteresis loops at temperatures of 1.8 K, 50 K, and 130 K. The coercivity (Hc) varies with temperature and reaches 1060 Oe at 1.8 K, 4132 Oe at 50 K, 2116 Oe at 100 K, 1402 Oe at 130 K and dropps to 11 Oe at 180 K. Therefore, in the observed magnetic transition at 128 K (see Fig. 5), as evidenced by the broad feature in magnetic susceptibility, a nonzero coercivity is discernible above 128 K, particularly at 130 K, and the phenomenon only disappears only at temperatures close to 180 K. Furthermore, the remanence magnetization is strongly temperature-dependent, with values of 12 emu/M at 1.8 K, 24 emu/mol at 50 K, 22 emu/mol at 100 K, and approximately 10 emu/mol at 130 K. Beyond the last transition point observed in the M(T) dependence of the sample (168 K for $AgF_2$ traces), the M(H) curve becomes linear without coercivity (measurements at 180 K and 250 K, see also ESI). These results agree with the magnetic susceptibility measurements and indicate a weak ferromagnetic character of the sample at low temperatures. However, since magnetic saturation was not reached even at lowest applied temperature under the maximum applied magnetic field of 70kOe (see ESI), this indicates the spin canting origin of ferromagnetism (as in $AgF_2$).

The magnetic behavior of $AgCoF_5$ may be related to that of structurally analogous compounds. Several mixed-valent paramagnetic transition metal pentafluorides with the *C*2/*c* structure have been documented[1,2,9] (see Supporting Information). Homometallic compounds such as $Mn_2F_5$ and $Cr_2F_5$ have been shown to be antiferromagnets by magnetic susceptibility measurements below 53 K[2] and 30 K[1], respectively. For the heteroatomic compound $MnCrF_5$ an antiferromagnetic order below 6 K was observed[17]. However, for $CrTiF_5$ and $CrVF_5$, the ferrimagnetism was observed below 26 K[1] and 40 K[1], respectively. This was attributed to the $3d^1$ and $3d^2$ configurations for $Ti^{III}$ and $V^{III}$ ions and to the simultaneous $3d^4$ configuration of $Cr^{II}$. Therefore, the different electronic configurations of $3d^6$ HS-$Co^{III}$ and $4d^9$ $Ag^{II}$, suggest the possibility of ferrimagnetism of $AgCoF_5$.

**Table. 2** Comparison of magnetic behavior and transition temperature for compounds with C2/c symmetry and mixed-valence transition metals.

| Compound | $T_t$ [K] | Magnetic nature/GS* |
|---|---|---|
| $Mn_2F_5$[2] | 53.4 | AF |
| $Cr_2F_5$[1] | 40 | AF |
| $CrTiF_5$[1] | 26 | Fi |
| $CrVF_5$[1] | 40 | Fi |
| $MnCrF_5$[17] | 6 | AF |
| $CdMnF_5$[9] | n.d. | n.d. |
| $AgCoF_5$# | $T_N$ = 128, $T_2$ = 3-9& | Complex behavior (most likely Fi) |

* AF – antiferromagnetic, Fi – ferrimagnetic, # - this work, & - depending on the field applied (ESI)

The complex magnetic behavior of the $AgCoF_5$ compound necessitates a more comprehensive characterization. Techniques such as muon spin resonance spectroscopy or neutron diffraction, supported by quantum mechanical calculations in a non-collinear model and with explicit inclusion of spin-orbit coupling, should be employed. While these measurements and calculations remain to be performed in the future, we have conducted a preliminary analysis of the magnetic superexchange (SE) constants.

The exchange coupling constants were calculated using the Heisenberg Hamiltonian formalism based on the energy calculations of the respective spin configurations (see ESI) within the broken symmetry method. We have determined the five closest metal contacts – both between silver and cobalt ($J^b_1$, $J^a_2$, $J^c_4$), as well as the homoatomic ones ($J^c_3$, $J_5$). A detailed description of the SE paths is shown presented in Figure SI7. The determined values of the superexchange (SE) constants obtained from DFT+U calculations (for $U_{Ag}$= 5 and 8 eV), SCAN and HSE06 methods are listed in the table below. A negative value indicates antiferromagnetic ordering between the spins, while a positive values indicate ferromagnetic ordering. In the context of magnetic interaction strength, we mainly refer to the absolute values of the particular constant describing the interactions, denoted as |J|.

Analyzing the differences in the calculated values for relevant superexchange constants between diverse methods, the largest discrepancies are observed for $J^b_{1(Co-Ag)}$. This situation is likely not only a result of differences in the exchange-correlation functionals employed, but also stems from the variations in the geometry of the systems obtained during optimization by each method – specifically, the distances between the paramagnetic centers and the angle of the bond formed through fluorine. All these parameters together with the SE constants, are listed in Table 3 for all the methods used. Recently, however, we have shown that among various exchange-correlation (XC) functionals, the SCAN method provides the best agreement between experimental and theoretical $J_{2D}$ value for $AgF_2$, with an error margin of only 4%[33]. Therefore, we assume here -62 meV as a reliable value for $J^b_{1(Co-Ag)}$ and restrict ourselves to the SCAN results in the following analysis.

**Table. 3** The values of the SE constants [in meV] determined by the DFT+U, SCAN and HSE06 methods, together with the directions, angles and distances of the superexchange pathways between the metal sites.

| Direction | Parameter | DFT+U ($U_{Ag}$=5 eV) | DFT+U ($U_{Ag}$=8 eV) | SCAN | HSE06 |
|---|---|---|---|---|---|
| [010] | $J^b_{1(Co-Ag)}$ [meV] | -47.7 | -39.3 | -62.0 | -39.5 |
|  | d [Å] | 3.822 | 3.803 | 3.874 | 3.831 |
|  | angle | 158.9° | 158.6° | 164.1° | 159.6° |
| [100] | $J^a_{2(Co-Ag)}$ [meV] | -6.5 | -6.5 | -4.8 | -6.7 |
|  | d [Å] | 3.593 | 3.579 | 3.635 | 3.600 |
|  | angle | 127.9° | 127.7° | 130.1° | 129.1° |
| [001] | $J^c_{3(Co-Co)}$ [meV] | -8.3 | -8.2 | -10.3 | -8.0 |
|  | d [Å] | 3.768 | 3.754 | 3.793 | 3.763 |
|  | angle | 158.9° | 158.9° | 162.4° | 160.2° |
| [001] | $J^c_{4(Ag-Ag)}$ [meV] | -1.1 | -0.4 | -1.3 | +0.7 |
|  | d [Å] | 3.768 | 3.754 | 3.793 | 3.763 |
|  | angle | 107.9° | 108.4° | 109.1° | 106.9° |
| [101] | $J_{5(Ag-Co)}$ [meV] | -1.2 | -1.4 | -1.9 | -0.5 |
|  | d [Å] | 3.961 | 3.943 | 3.915 | 3.967 |
|  | angle | 121.3 | 121.1 | 120.7 | 121.3 |

According to all the theoretical methods used, the ground state of $AgCoF_5$ is antiferromagnetic with a quasi-G-type magnetic structure (as illustrated in the Figure 9), which means that the antiferromagnetic exchange occurs between all six corner-sharing metal centers. In addition, $Cr_2F_5$, which has the same crystal structure as the compound discussed in this work, has a magnetic structure identical to that determined here for $AgCoF_5$, as previously suggested[57]. However, the specific values of the superexchange constants (SE) for $Cr_2F_5$ have not yet been determined.

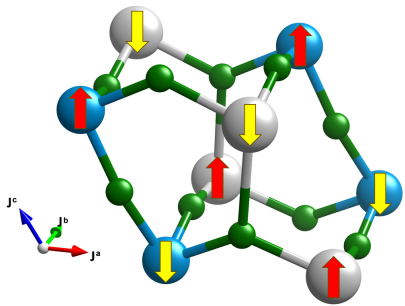

**Figure 9.** The G-type antiferromagnetic structure, calculated as the magnetic ground state for $AgCoF_5$. A comprehensive description of the superexchange paths and all spin configurations considered can be found in the Electronic Supplementary Information (ESI).

In the rectangular lattice along the *ab* plane, we identified two SE constants: $J^b_1$ and $J^a_2$, both for heteroatomic (Ag-Co) magnetic interactions. The most important SE constant, labelled as $J^b_{1(Co-Ag)}$, is related to the Co-Ag interaction along the b-axis [010] via an F bridge. Its value from the SCAN method is -62.02 meV (Table 3). Since the Co-F-Ag bond angle along [010] is 164.1° (SCAN), the observation of a relatively strong antiferromagnetic exchange along this SE path is rather expected in line with the GKA (Goodenough-Kanamori-Anderson) rules [58–60]. This interaction encompasses the $e_g$ electrons of Co s=4/2 (high-spin) and Ag s=1/2 cations, since the magnetic moments for these paramagnetic sites have been computed as 3.22 $\mu_B$ and 0.58 $\mu_B$ (DFT+U), respectively (as for typical HS-$Co^{3+}$[28] and $d^9$ $Ag^{2+}$[33,61], ESI). The subsequent SE constant in the rectangular lattice within the (001) plane, which describes the strength of the Co-Ag interaction along the [100] direction ($J^a_{2(Co-Ag)}$), appears to be considerably weaker than the dominant interaction. The SCAN method gives the value of -4.8 meV. The Co-F-Ag angle along the a-axis is 130.1° (SCAN) and thus deviates significaly from 180°, which considerably weakens the antiferromagnetic interaction.

For Co-F-Co chains along [001], the superexchange is stronger than for Co-F-Ag (-10.3 meV). The other two exchange constants, $J^c_{4(Ag-Ag)}$ and $J^5_{(Ag-Co)}$, are significantly weaker and show negative values. Our results are similar to those for isostructural $Cr_2F_5$, where Monte Carlo simulations suggested that the three strongest superexchange (SE) constants should be negative, indicating of antiferromagnetic ordering between the spins, while the other two constants are significantly weaker[57].

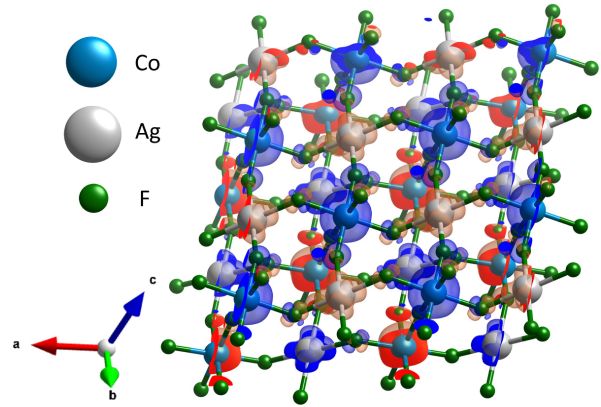

**Figure 10.** Figure presents part of $AgCoF_5$ 2x2x1 supercell. The spin density localized at the fluorine atoms, illustrates their pivotal role as intermediaries (bridge) in the superexchange interaction mechanism. The spin density was calculated using Density Functional Theory augmented with Hubbard U ($U_{Ag}$ = 5 eV), and is shown with an isosurface threshold set at 0.008 electrons per Å$^3$, where blue and red refer to alpha and beta spin density excess.

The study of the SE constants in $AgCoF_5$ clearly shows a pronounced anisotropy of these interactions. It was demonstrated that the interaction characterized by $J^b_{1(Ag-Co)}$ is considerably more robust than the other two, which are described as $J^a_{2(Co-Ag)}$ and $J^c_{3(Co-Co)}$. To assess the extent of this anisotropy, one can calculate the ratio J'/J'', where J'' represents the constant with the highest absolute value – in this case $J^b_{1(Ag-Co)}$, and J' as that of $J^a_{2(Co-Ag)}$ or $J^c_{3(Co-Co)}$. The most significant anisotropy is evident in the SCAN results. Considering the constant J' as the constant describing the interaction between S=4/2 ($Co^{III}$) and S=1/2 ($Ag^{II}$), the ratio is 7.69*10$^{-2}$ (J'= $J^a_{2(Co-Ag)}$). For the SE constant, which describes the interaction between two cobalt ions (both S=4/2), the ratio is 1.66*10$^{-1}$ (J'= J$^c_{3(Co-Co)}$). The results for all methods are shown in the Table 4.

The J'/J'' ratios calculated in our study are significantly higher than those typically found in nearly ideal 1D antiferromagnets. For instance, in FeF$_3$(4,4'-bpy), where superexchange (SE) occurs along the Fe-F-Fe (S=5/2) chains, the J'/J'' ratio is less than 3.2 · 10$^{-5}$[62]. This tendency can also be observed for ternary compounds with Ag$^{II}$ and F, such as KAgF$_3$ (2.1·10$^{-2}$)[54]. Additionally, a similar ratio was also found for the quasi-1D magnet Bi$_2$Fe(SeO$_3$)$_2$OCl$_3$, where J'/J'' is equal to 8.0·10$^{-2}$[63]. It is important to note that in all these compounds the chains are either completely or quasi-isolated, in contrast to AgCoF$_5$, which has a 3D network of metal-ligand-metal bonds.

**Table. 4** The anisotropy ratio J'/J'', regarding J' = J$^a_{2(Co-Ag)}$ or J$^c_{3(Co-Co)}$ and J'' as J$^b_{1(Co-Ag)}$.

| Method | J'/J'' ratio | |
|---|---|---|
| | J' = J$^a_{2(Co-Ag)}$ | J' = J$^c_{3(Co-Co)}$ |
| **DFT+U (U$_{Ag}$=5 eV)** | 1.37·10$^{-1}$ | 1.73·10$^{-1}$ |
| **DFT+U (U$_{Ag}$=8 eV)** | 1.66·10$^{-1}$ | 2.09·10$^{-1}$ |
| **SCAN** | 7.69·10$^{-2}$ | 1.66·10$^{-1}$ |
| **HSE06** | 1.71·10$^{-1}$ | 2.02·10$^{-1}$ |

All this together suggests that AgCoF$_5$ does not exhibit characteristics of a canonical one-dimensional (1D) antiferromagnet, but rather should be classified as a quasi-three-dimensional (3D) antiferromagnet with significant magnetic anisotropy within the rectangular mixed spin lattice (along *ab*) (J' and J'' < J$_{1D}$ by around 10$^{-1}$). In view of this, one can tentatively assign the observed transition at 128K to the onset of antiparallel alignment of the Co and Ag spins along the crystallographic b-axis.

Analysis of spin density for the ground state magnetic solution (Figure 10) shows that there is substantial spin polarization on F atoms, which constitute key intermediaries of the superexchange.

**Electronic properties of AgCoF$_5$**

The inspection of the electronic density of states indicates the presence of an energy gap between occupied and unoccupied states, ranging from 1.382 eV (DFT+U) to 2.476 eV (HSE06). The nature of this gap can be described as charge-transfer (CT) according to the ZSA (Zaanen-Sawatzky-Allen) classification[64]. The main contribution to the valence band just below the Fermi level (zero on the x-axis of Fig. 11) originates from the *p* states of fluorine, while, the conduction band consists mainly of the unoccupied states of the metals – silver and/or cobalt. In DFT+U (U$_{Ag}$ = 5 eV) and HSE06 calculations, the electronic states proximal to the Fermi level (0 at the Energy axis in Fig. 9) within the conduction band predominantly consist of Ag$_{(dx2-y2)}$ states. In addition, these states also provide a substantial contribution to the valence states near the Fermi level. A comparable observation is noted for Co$_{(dx2-y2)}$ and Co$_{(dxy)}$ states, where both valence and conduction bands, close to the Fermi level, have the greatest contribution from those particular states, across all Co *d* orbitals. It is important to note that the unoccupied states of silver and cobalt are different, each forming relatively narrow bands (HSE06 and DFT+U, where U$_{Ag}$ = 5 eV). In the unoccupied band, alongside the metal states, F$_p$ states are also present, indication a notable hybridization between the ligand (F$_p$) and metal (Ag$_d$ and/or Co$_d$) states. This also suggests the existence of substantial amount of 'holes' in the fluorine orbitals. Similarly, the coexistence of ligand and metal states in the same energy range from -8 to 0 eV below the Fermi level, implies significant orbital hybridization. This covalence of the chemical bonds fits well with the relatively high values of the superexchange (SE) constants, particularly for J$^b_{1(Co-Ag)}$, where the F*p* orbitals mediate the magnetic interaction.

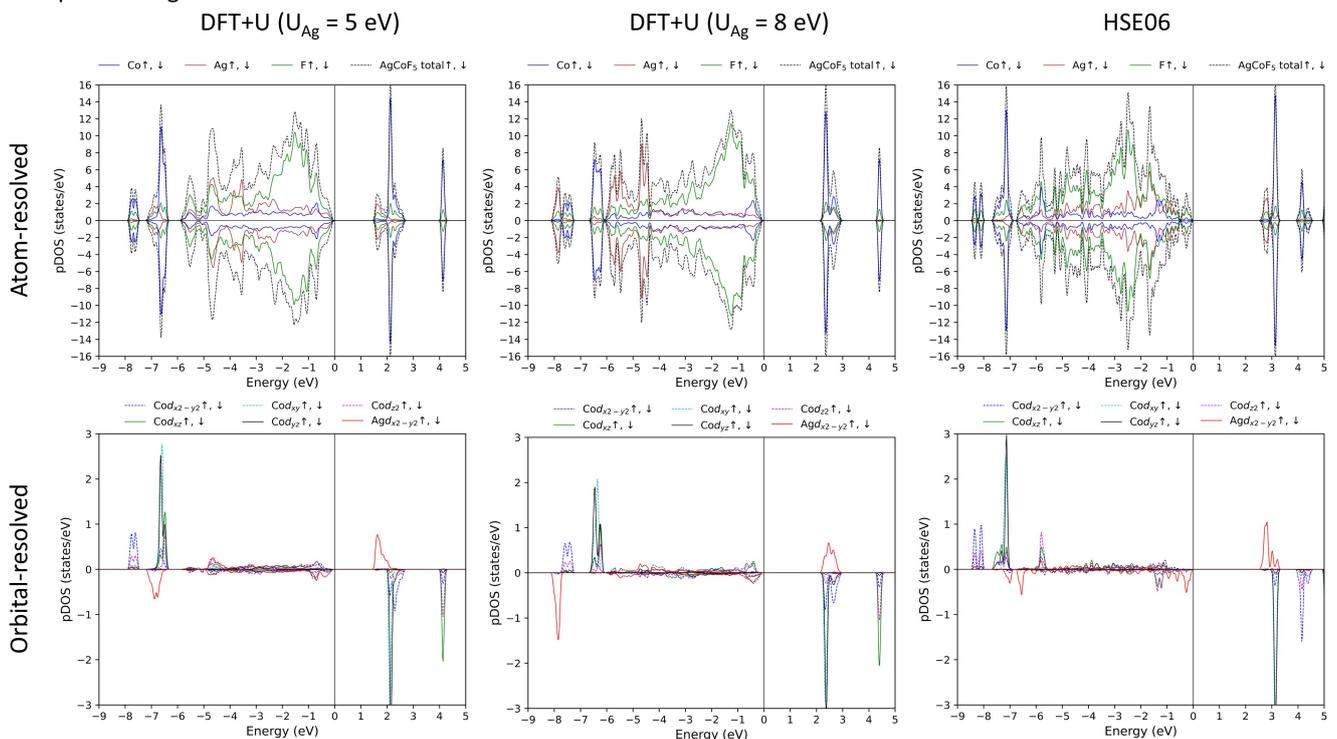

**Figure 11.** Density of states for AgCoF$_5$ calculated with the DFT+U method (U$_{Co}$ is equal to 5 eV). For orbital resolved DOS only the *d* states of Co are shown, for Ag – only d$_{(x2-y2)}$.

**Table. 5** The energy of formation for Ag$^{II}$M$^{III}$F$_5$ compounds according to eq3. The results were obtained using DFT+U method.

| M$^{III}$ | dG [eV/mol] | dE [kJ/mol] | dV [Å$^3$] | Mag. mom [μB] | |
| --- | --- | --- | --- | --- | --- |
| | | | | Ag | M$^{III}$ |
| Ag | -0.09 | -8.21 | 6.24 | +/-0.82 | +/-0.49 |
| Cu* | -0.06 | -5.97 | 5.97 | +/-0.54 | +/-1.06 |
| Co | -0.03 | -3.25 | 5.25 | +/-0.58 | +/-3.22 |
| Fe | 0.06 | 5.83 | 4.71 | +/-0.59 | +/-4.32 |
| Ni | 0.06 | 6.07 | 4.36 | +/-0.65 | +/-1.01 |
| Ga | 0.12 | 11.89 | 4.96 | +0.51 | +/-0.01 |
| Al | 0.21 | 20.28 | 2.92 | +/-0.65 | +/-0.03 |
| Sc | 0.26 | 24.63 | 2.00 | +/-0.51 | +/-0.03 |
| Au | 0.26 | 25.40 | 3.08 | +/-0.56 | +/-1.01 |

* structure of CuF$_3$ not known experimentally but provisionally assumed to be isostructural with that of FeF$_3$, as suggested by recent calculations[4]

**Theoretical insight into the stability of the different members of the AgMF$_5$ family**

Encouraged by the preparation of AgCoF$_5$, we decided to theoretically investigate the stability of this and several other members of the AgMF$_5$ family with trivalent M cations. It turned out that the formation of AgCoF$_5$ from AgF$_2$ and CoF$_3$ is slightly exothermic, about -3 kJ/mol (Table 5). Therefore, we have screened the following reactions:

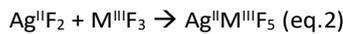

Ag$^{II}$F$_2$ + M$^{III}$F$_3$ → Ag$^{II}$M$^{III}$F$_5$ (eq.2)

and we have calculated:

$$dG = G_{AgMF_5} - G_{AgF_2} - G_{MF_3} \quad eq.3$$
$$dV = V_{AgMF_5} - V_{AgF_2} - V_{MF_3} \quad eq.4$$

We found that only in three cases, the enthalpy (eq3) of bimetallic pentafluoride formation was negative: for the already mentioned AgCoF$_5$ and for two hypothetical phases – AgCuF$_5$ and Ag$^{II}$Ag$^{III}$F$_5$. In the case of AgCuF$_5$, for calculations of eq.3 and eq.4 we adapted theoretically predicted CuF$_3$ structure[4]. Moreover, the latest experimental approach in the HT solid state reaction of CuF$_2$/AgF$_2$ mixtures with F$_2$ overpressure revealed in the formation of Ag$^{II}$/Cu$^{II}$-F solid solutions without oxidation to Cu$^{III}$ fluoride[24]. For these reasons, the significance of this result is somewhat diminished.

However, the case of Ag$^{II}$Ag$^{III}$F$_5$ is exceptionally interesting for three reasons: (i) it contains two powerful oxidizers, Ag$^{II}$ and Ag$^{III}$, simultaneously in its structure; (ii) it has two paramagnetic centers – Ag$^{2+}$(d$^9$) and HS-Ag$^{3+}$(d$^8$) and (iii) a compound with this stoichiometry has already been prepared. However, the structure produced is not monoclinic but triclinic and it contains LS-Ag$^{III}$[3].

This raises the question of the relative energetic stability of these two polytypes – the hypothetical monoclinic C2/c and the synthesized triclinic P-1. Our preliminary DFT+U calculations indicate that the C2/c structure is slightly more energetically favorable by 21 meV/FU (including Zero-Point Energy, ZPE) compared to the (P-1) structure type. Moreover, the most significant superexchange (SE) constants between s=1/2 (d$^9$) and s=1 (d$^8$) silver sites (J$_{2D}$) in the ab plane, calculated for the monoclinic Ag$^{II}$Ag$^{III}$F$_5$ structure are at -100 meV and for two Ag$^{III}$ sites around -93 meV (DFT+U, U$_{Ag}$ = 5 eV). A further description of this new hypothetical polymorphic form of a known compound is beyond the scope of this paper and will be explored elsewhere. It suffices to say that the theory suggests the existence of an Ag$^{II}$Ag$^{III}$F$_5$ phase with HS-Ag$^{III}$ that is analogous to the prepared Ag$^{II}$Co$^{III}$F$_5$.

Finally, we experimentally verified the reactions between NiF$_2$, AlF$_3$, and FeF$_3$ separately with AgF$_2$. Under conditions coresponding to those of the successful synthesis of AgCoF$_5$, no formation of new phases was observed – the powder contained only the unreacted substrates after annealing at high temperature. Therefore, the theory seems to accurately describe the lack of energetic preference for the formation of AgNiF$_5$ (starting from NiF$_3$), AgAlF$_5$ and AgFeF$_5$.

## Conclusions and prospects

We have prepared a new member of the family of bimetallic pentafluorides, albeit the first to contain Ag$^{II}$. The monoclinic AgCoF$_5$ crystallizes in the monoclinic unit cell. Apart from the good Rietveld fit parameters, we have confirmed the structure by phonon calculations, with very good agreement between the computed and observed bands in the IR and Raman spectra. The title compound shows a complex magnetic structure that can tentatively characterized as ferrimagnetic. This magnetic behavior results from the d$^9$ silver and the high-spin cobalt – d$^6$, both of which have open d shells. The quasi-layered structural configuration of the compound facilitates antiferromagnetic interactions between disparate spin states, in particular S=4/2 (Co$^{III}$, d$^6$ high spin), and S=1/2 (Ag$^{II}$, d$^9$). These interactions are favored by a superexchange mechanism mediated by a fluorine bridge, with the appreciable interaction strength estimated at -62 meV (SCAN), along crystallographic b axis. The next two superexchange interactions are much weaker, with magnetic anisotropy on the order of 10$^{-1}$.

For an accurate description of the magnetic properties of the compound, the application of more sophisticated methods, such as muon and neutron techniques, would be essential. In addition, obtaining the compound in the form of single crystals would significantly improve the possibilities of determining the magnetic structure of the titled compound, especially for measurements related to specific crystallographic axes. Experiments to grow single crystals will be carried out in the future.

Furthermore, based on theoretical calculations, we have indicated the possibility of the existence of a mixed-valence compound Ag$^{II}$Ag$^{III}$F$_5$ with HS-Ag$^{III}$ (C2/c). Here, the calculated predominant SE contant $J_{2D}$ value is -93 meV (DFT+U). It is worthwhile to investigate this fascinating system by experiments.

## Acknowledgements

W.G. is grateful to the Polish National Science Center (NCN) for project Maestro (2017/26/A/ST5/00570). Research was carried out with the use of CePT infrastructure financed by the European Union – the European Regional Development Fund within the Operational Programme "Innovative economy" for 2007-2013 (POIG.02.02.00-14-024/08-00). ZM acknowledges the financial support of the Slovenian Research and Innovation Agency (research core funding No. P1-0045; Inorganic Chemistry and Technology). Quantum mechanical calculations were performed using supercomputing resources of the ICM UW (project SAPPHIRE [GA83-34]). The authors are grateful to prof. Jose Lorenzana (CNR, Rome) for valuable remarks to this work.

## Methods

**Synthetic details**

The synthesis method applied was the same as in previous studies on $AgF_2/CuF_2$ systems[24]. Further details can be found in section 1.1 of the paper.

**Magnetic measurements**

The powder samples, weighing about 35 mg, were encapsulated in Teflon tubes under a dry argon atmosphere before being transferred to the measurement chamber. Magnetic characterizations were conducted using a Quantum Design MPMS SQUID VSM magnetometer. Magnetic susceptibility as a function of temperature was recorded in the range of 2 to 300 K under different applied magnetic fields under both field-cooled (FC) and zero-field-cooled (ZFC) conditions with a temperature increment of 1K. On the same sample M(H) dependence were recorded. The data acquired were adjusted for the baseline contribution of an empty sample holder.

**XRD measurements**

The powder samples were enclosed in perfluorinated quartz capillaries with a diameter of 0.5 mm under a dry argon atmosphere. The experiments were performed using a PANalytical X'Pert Pro diffractometer equipped with a linear PIXcel Medipix2 detector, with a parallel beam of $CoK_{\alpha1}$ and $CoK_{\alpha2}$ radiation with an intensity ratio of 2:1, at ambient temperature. The Pseudo-Voight function was utilized for peak shape analysis, while the Berar-Baldinozzi function was used to analyze the peak asymmetry. The background signals were modeled with 20-36 Legendre polynomials. All analyses included absorption correction ($U_{ISO}$) as implemented in the software Jana2006[39].

**IR and RAMAN measurements**

For the RAMAN measurements powder samples were enclosed in perfluorinated quartz capillaries with a diameter of 0.5 mm under a dry argon atmosphere. The measurements were carried out on T64000 spectrometer from Horiba–Jobin Yvon equipped with a liquid nitrogen cooled Si detector with the Mitutoyo long-working distance lenses. The excitation line with a wavelength of 532 nm was used. We collected signals from different locations to identify possible decomposition features and exclude them from the analysis. To avoid photodecomposition of the sample, we typically conducted 420 accumulations, with each lasting 40 seconds, using a beam power of between 0.3 and 3 mW.

The IR spectra were recorded with a Vertex 80v spectrometer from Bruker. A small amount of the powder was placed under dry argon between the HDPE windows and tightly closed in the measurement cell.

**Computational details**

The computations were conducted utilizing the Density Functional Theory (DFT) framework as implemented in the VASP 5.4.4 software[40]. A Generalized Gradient Approximation (GGA) type PBEsol functional[41]was employed, coupled with the projected augmented wave method[42,43]. To account for the on-site Coulombic interactions of the d electrons, Hubbard ($U_d$) and Hund ($J_H$) parameters were introduced, following the DFT+U formalism as suggested by Liechtenstein[44]. In the DFT+U and the $J_H$ parameter was set to 1 eV[45], while the $U_d$ values for Ag were set to 8 or 5 eV and for Co as 5 eV, respectively[29,46]. Other methods used the SCAN[47] and HSE06[48] approaches. A plane-wave cutoff energy of 520 eV was utilized for all systems. The k-spacing parameters were established at 0.032 Å$^{-1}$ (0.048 Å$^{-1}$) for geometry optimization and 0.022 Å$^{-1}$ (0.032 Å$^{-1}$) for achieving self-consistent-field convergence in the DFT+U and SCAN methods (in arrays values for HSE06). Convergence criteria of 10$^{-9}$ eV for electronic steps and 10$^{-7}$ for ionic steps were applied.

**Graphical presentation**

All figures of the structures were visualized with the VESTA software[49].

## Notes and references

‡ Crystal structure of $AgCoF_5$ has been deposited into ICSD database (No. 2332022).

§ Electronic supplementary information (ESI) available.

## References

1. A. Tressaud, J. M. Dance, J. Ravez, J. Portier, P. Hagenmuller, J. B. Goodenough, *Materials Research Bulletin* **1973**, *8*, 1467–1477.
2. J. Bandemehr, F. Zimmerhofer, S. I. Ivlev, C. Pietzonka, K. Eklund, A. J. Karttunen, H. Huppertz, F. Kraus, *Inorg. Chem.* **2021**, *60*, 12651–12663.
3. R. Fischer, B. G. Müller, *Z. anorg. allg. Chem.* **2002**, *628*, 2592–2596.
4. N. Rybin, D. Y. Novoselov, D. M. Korotin, A. R. Oganov, *Phys. Chem. Chem. Phys.* **2021**, *23*, 15989–15993.
5. M. Tramšek, B. Žemva, *Acta Chim. Slov.* **2002**, *49*, 209–220.
6. J. V. Rau, V. R. Albertini, N. S. Chilingarov, S. Colonna, M. D. Michiel, *Chem. Lett.* **2002**, *31*, 664–665.
7. J. V. Rau, V. Rossi Albertini, N. S. Chilingarov, M. Di Michiel, S. Colonna, I. N. Ioffe, L. N. Sidorov, *BCSJ* **2003**, *76*, 1165–1169.

8   H. J. Seifert, H. W. Loh, K. Jungnickel, *Zeitschrift anorg allge chemie* **1968**, *360*, 62–69.
9   U. Müller, R. Hoppe, *Z. Anorg. Allg. Chem.* **1990**, *583*, 205–208.
10  J. Graulich, W. Massa, D. Babel, *Z. anorg. allg. Chem.* **2003**, *629*, 365–367.
11  S. M. Eicher, J. E. Greedan, *Journal of Solid State Chemistry* **1984**, *52*, 12–21.
12  [12]   R. Von Der Mühll, S. Andersson, J. Galy, *Acta Crystallogr B Struct Crystallogr Cryst Chem* **1971**, *27*, 2345–2353.
13  R. Georges, J. Ravez, R. Olazcuaga, P. Hagenmuller, *Journal of Solid State Chemistry* **1974**, *9*, 1–5.
14  R. von der Mühll, F. Daut, J. Ravez, *Journal of Solid State Chemistry France* **1973**, *8*, 206–212.
15  A. Tressaud, J. M. Parenteau, J. M. Dance, J. Portier, P. Hagenmuller, **1973**, *8*.
16  M. Body, G. Silly, C. Legein, J.-Y. Buzaré, F. Calvayrac, P. Blaha, *Journal of Solid State Chemistry* **2005**, *178*, 3655–3661.
17  G. Ferey, R. Pape, *Acta Cryst.* **1978**, 1084–1091.
18  W. J. Casteel, G. Lucier, R. Hagiwara, H. Borrmann, N. Bartlett, *Journal of Solid State Chemistry* **1992**, *96*, 84–96.
19  R. Hoppe, G. Siebert, *Z. Anorg. Allg. Chem.* **1970**, *376*, 261–267.
20  B. Müller, R. Hoppe, *Z. Anorg. Allg. Chem.* **1972**, *392*, 37–41.
21  [21]   G. Lucier, J. Muenzenberg, W. J. Casteel, N. Bartlett, *Inorg. Chem.* **1995**, *34*, 2692–2698.
22  M. A. Domański, W. Grochala, *Zeitschrift für Naturforschung B* **2021**, *76*, 751–758.
23  M. Domański, J. van Leusen, M. Metzelaars, P. Kögerler, W. Grochala, *J. Phys. Chem. A* **2022**, *126*, 9618–9626.
24  D. Jezierski, K. Koteras, M. Domański, P. Połczyński, Z. Mazej, J. Lorenzana, W. Grochala, *Chemistry – A European Journal* **2023**, *29*, e202301092.
25  P. Fischer, G. Roult, D. Schwarzenbach, *Journal of Physics and Chemistry of Solids* **1971**, *32*, 1641–1647.
26  E. O. Wollan, H. R. Child, W. C. Koehler, M. K. Wilkinson, *Phys. Rev.* **1958**, *112*, 1132–1136.
27  J. Gawraczyński, D. Kurzydłowski, R. A. Ewings, S. Bandaru, W. Gadomski, Z. Mazej, G. Ruani, I. Bergenti, T. Jaroń, A. Ozarowski, S. Hill, P. J. Leszczyński, K. Tokár, M. Derzsi, P. Barone, K. Wohlfeld, J. Lorenzana, W. Grochala, *Proc Natl Acad Sci USA* **2019**, *116*, 1495–1500.
28  S. Mattsson, B. Paulus, *Journal of Computational Chemistry* **2019**, *40*, 1190–1197.
29  R. Piombo, D. Jezierski, H. P. Martins, T. Jaroń, M. N. Gastiasoro, P. Barone, K. Tokár, P. Piekarz, M. Derzsi, Z. Mazej, M. Abbate, W. Grochala, J. Lorenzana, *Phys. Rev. B* **2022**, *106*, 035142.
30  N. Bachar, K. Koteras, J. Gawraczynski, W. Trzciński, J. Paszula, R. Piombo, P. Barone, Z. Mazej, G. Ghiringhelli, A. Nag, K.-J. Zhou, J. Lorenzana, D. Van Der Marel, W. Grochala, *Phys. Rev. Research* **2022**, *4*, 023108.
31  S. Bandaru, M. Derzsi, A. Grzelak, J. Lorenzana, W. Grochala, *Phys. Rev. Materials* **2021**, *5*, 064801.
32  A. Grzelak, H. Su, X. Yang, D. Kurzydłowski, J. Lorenzana, W. Grochala, *Phys. Rev. Materials* **2020**, *4*, 084405.
33  [33]   D. Jezierski, W. Grochala, *submitted to Physical Review Materials* **2023**.
34  W. Grochala, R. Hoffmann, *Angew. Chem. Int. Ed.* **2001**, *40*, 2742–2781.
35  A. Grzelak, M. Derzsi, W. Grochala, *Inorg. Chem.* **2021**, *60*, 1561–1570.
36  D. Jezierski, A. Grzelak, X. Liu, S. K. Pandey, M. N. Gastiasoro, J. Lorenzana, J. Feng, W. Grochala, *Phys. Chem. Chem. Phys.* **2022**, *24*, 15705–15717.
37  M. A. Domański, M. Derzsi, W. Grochala, *RSC Adv.* **2021**, *11*, 25801–25810.
38  A. Grzelak, J. Lorenzana, W. Grochala, *Angew. Chem. Int. Ed.* **2021**, *60*, 13892–13895.
39  V. Petříček, M. Dušek, L. Palatinus, *Zeitschrift für Kristallographie - Crystalline Materials* **2014**, *229*, 345–352.
40  G. Kresse, J. Furthmüller, *Phys. Rev. B* **1996**, *54*, 11169–11186.
41  J. P. Perdew, A. Ruzsinszky, G. I. Csonka, O. A. Vydrov, G. E. Scuseria, L. A. Constantin, X. Zhou, K. Burke, *Phys. Rev. Lett.* **2008**, *100*, 136406.
42  P. E. Blöchl, *Phys. Rev. B* **1994**, *50*, 17953–17979.
43  G. Kresse, D. Joubert, *Phys. Rev. B* **1999**, *59*, 1758–1775.
44  A. I. Liechtenstein, V. I. Anisimov, J. Zaanen, *Phys. Rev. B* **1995**, *52*, R5467–R5470.
45  V. I. Anisimov, J. Zaanen, O. K. Andersen, *Phys. Rev. B* **1991**, *44*, 943–954.
46  F. Zhou, M. Cococcioni, C. A. Marianetti, D. Morgan, G. Ceder, *Phys. Rev. B* **2004**, *70*, 235121.
47  J. Sun, A. Ruzsinszky, J. P. Perdew, *Phys. Rev. Lett.* **2015**, *115*, 036402.
48  A. V. Krukau, O. A. Vydrov, A. F. Izmaylov, G. E. Scuseria, *The Journal of Chemical Physics* **2006**, *125*, 224106.
49  K. Momma, F. Izumi, *J Appl Cryst* **2008**, *41*, 653–658.
50  W. Levason, C. A. McAuliffe, *Coordination Chemistry Reviews* **1974**, *12*, 151–184.
51  M. A. Neumann, *J Appl Cryst* **2003**, *36*, 356–365.
52  D. T. Durig, J. R. Durig, in *Low Temperature Molecular Spectroscopy* (Ed.: R. Fausto), Springer Netherlands, Dordrecht, **1996**, pp. 477–504.
53  J. Fábry, R. Krupková, *Ferroelectrics* **2008**, *375*, 59–73.
54  D. Kurzydłowski, Z. Mazej, Z. Jagličić, Y. Filinchuk, W. Grochala, *Chem. Commun.* **2013**, *49*, 6262.
55  I. Dzyaloshinsky, *Journal of Physics and Chemistry of Solids* **1958**, *4*, 241–255.
56  T. Moriya, *Phys. Rev.* **1960**, *120*, 91–98.
57  P. Lacorre, G. Ferey, J. Pannetier, *Journal of Solid State Chemistry* **1992**, *96*, 227–236.
58  J. B. Goodenough, *Phys. Rev.* **1955**, *100*, 564–573.
59  J. Kanamori, *Journal of Physics and Chemistry of Solids* **1959**, *10*, 87–98.
60  P. W. Anderson, *Phys. Rev.* **1950**, *79*, 350–356.
61  C. Miller, A. S. Botana, *Phys. Rev. B* **2020**, *101*, 195116.
62  H. Lu, T. Yamamoto, W. Yoshimune, N. Hayashi, Y. Kobayashi, Y. Ajiro, H. Kageyama, *J. Am. Chem. Soc.* **2015**, *137*, 9804–9807.
63  P. S. Berdonosov, E. S. Kuznetsova, V. A. Dolgikh, A. V. Sobolev, I. A. Presniakov, A. V. Olenev, B. Rahaman, T. Saha-Dasgupta, K. V. Zakharov, E. A. Zvereva, O. S. Volkova, A. N. Vasiliev, *Inorg. Chem.* **2014**, *53*, 5830–5838.
64  H. Eskes, G. A. Sawatzky, *Phys. Rev. Lett.* **1988**, *61*, 1415–1418.

# SUPPLEMENTARY INFORMATION

*Novel $Ag^{II}Co^{III}F_5$ Ternary Fluoride: Synthesis, Structure and Magnetic Characteristics*

Daniel Jezierski[1]*, Zoran Mazej[2]*, Wojciech Grochala[1]*

**Contents**

SI.I Overview of the $A^{II}B^{III}F_5$ fluoride family

SI.II Structural data for $AgCoF_5$

SI.III Comparison of Rietveld fits for *C*2/*c* and *C*2/*m* space groups

SI.IV Magnetic data

SI.IV Experimental and theoretical phonon frequencies of $AgCoF_5$

SI.V Results of superexchange calculations

SI.VI References

# SI.I Overview of the $A^{II}B^{III}F_5$ fluoride family

| | Compound | Mangetic GS* | $T_t$ [K] | REF | STRUCTURE |
|---|---|---|---|---|---|
| TM-TM | $CrTiF_5$ | Fi | 26 | [1] | *C*2/*c* ($Cr_2F_5$- type) |
| | $CrVF_5$ | Fi | 40 | [1] | *C*2/*c* ($Cr_2F_5$- type) |
| | $MnCrF_5$ | AF | 6 | [2] | *C*2/*c* |
| | $CdMnF_5$ | n.d. | n.d. | [3] | *C*2/*c* |
| AE-TM | $CaFeF_5$ | AF | 21 | [4] | $P2_1/c$ |
| | $CaCrF_5$ | P | --- | [5] | *C*2/*c* |
| | $CaTiF_5$ | P | --- | [6] | *I*2/*c* |
| | $CaMnF_5$ | n.d. | n.d. | [3] | *C*2/*c* |
| | $BaFeF_5$ | AF | 35 | [7] | *I*4 |
| | $BaTiF_5$ | n.d. | n.d. | [6] | *I*4/*m* |
| | $BaVF_5$ | AF | 20 | [8] | *I*4 |
| | $SrVF_5$ | AF | 2 | [4] | $P2_1/c$ |
| | $SrCrF_5$ | P | --- | [8] | *I*4 |
| | $SrFeF_5$ | n.d. | n.d. | [9] | $P2_1/c$ |
| PT-TM | $CrAlF_5$ | P | --- | [1] | *C*2/*c* ($Cr_2F_5$- type) |
| | $MnAlF_5$ | P | --- | [10] | ORTORHOMBIC |
| AE-PT | alpha-$CaAlF_5$ | --- | --- | [5] | *C*2/*c* |
| | beta-$CaAlF_5$ | --- | --- | [11] | $P2_1/c$ |
| | $SrAlF_5$ | --- | --- | [7] | *I*4 |
| | $PbTF_5$ (T = Al, Ga) | --- | --- | [12] | n. d. |
| | $BaInF_5$ | --- | --- | [13] | n. d. |
| OTHER | $Mn(Al,Fe)F_5$ | Fi | 18-34 | [10] | ORT |

* Fi – ferrimagnetic, AF – antiferromagnetic, P – paramagnetic; n.d. stands for not determined

## SI.II Structural data for AgCoF$_5$

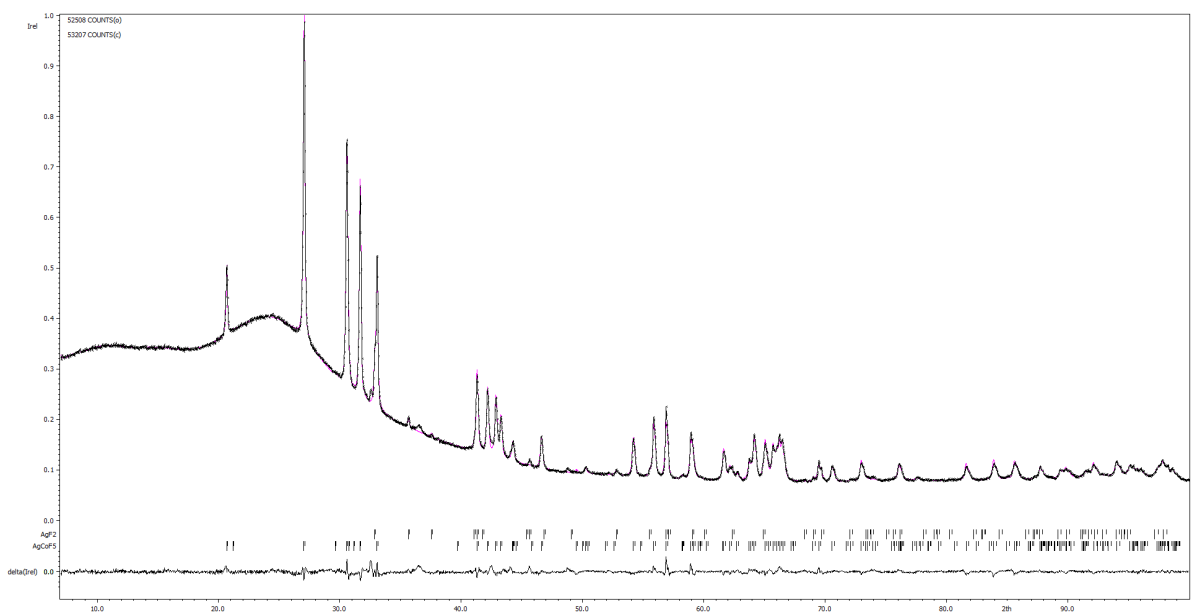

**Figure SI1.** Rietveld refinement of the X-ray pattern for the powder from the first synthetic approach. GoF = 1.15, Rp = 0.98, wRp = 1.36

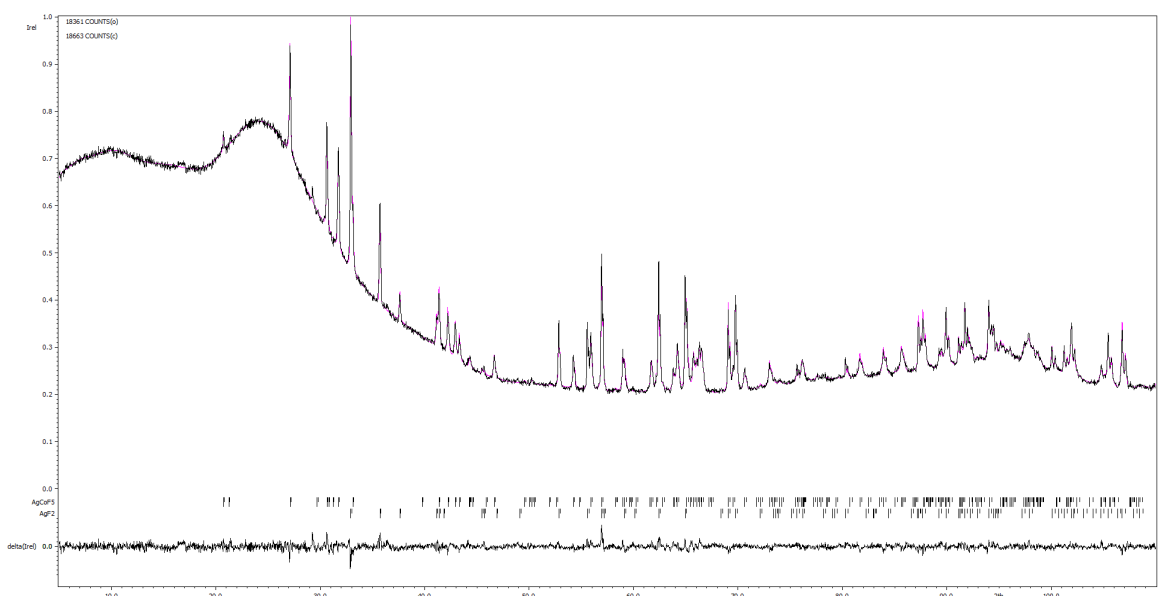

**Figure SI2.** Rietveld refinement of the X-ray pattern for the powder from the second synthetic approach. GoF = 1.66, Rp = 1.12, wRp = 1.68.

**Table SI1**. Unit cell parameters of AgF$_2$ and AgCoF$_5$ obtained from Rietveld refinement of two samples 1 (**S1**) and sample 2 (**S2**).

| S | Phase | Parameters | | | | | Molar ratio | Uncertainty [%] | Fitting parameters | | |
|---|---|---|---|---|---|---|---|---|---|---|---|
| | | a [Å] | b [Å] | c [Å] | V [Å$^3$] | β [°] | | | GoF | Rp | wRp |
| **1** | AgF$_2$ | 5.546 | 5.831 | 5.091 | 164.636 | 90 | 0.04 | 3 | 1.66 | 1.12 | 1.68 |
| | AgCoF$_5$ | 7.274 | 7.628 | 7.529 | 375.552 | 115.98 | 0.96 | 4 | | | |
| **2** | AgF$_2$ | 5.550 | 5.836 | 5.095 | 165.026 | 90 | 0.44 | 8 | 1.15 | 0.98 | 1.36 |
| | AgCoF$_5$ | 7.280 | 7.635 | 7.536 | 376.507 | 115.98 | 0.56 | 7 | | | |

**Table SI2**. Structural details of AgCoF$_5$ with atomic positions.

| Space group | C2/c (15) | Temperature | 298K | Radiation | Co Kα |
|---|---|---|---|---|---|
| Unit cell [Å] | \multicolumn{3}{l}{a = 7.274414 (2), b = 7.627744(2), c = 7.529471 (2)} | α = γ = 90° | β =115.976(4)° |
| **Atom** | **x** | **y** | **z** | **U$_{iso}$ [Å$^2$]** | **Occupancy** |
| Co | 0.000 | 0.000 | 0.000 | 0.003 (14) | 1 |
| Ag | 0.000 | 0.500 | 0.000 | 0.010 (13) | 1 |
| F1 | 0.000 | -0.038 (5) | 0.250 | 0.019 (18) | 1 |
| F2 | 0.793 (18) | 0.513 (6) | 0.624 (18) | 0.007 (15) | 1 |
| F3 | -0.028 (3) | 0.235 (2) | 0.448 (3) | 0.013 (15) | 1 |

It is important to note that the exact determination the positions of the fluorine atoms is a major challenge, even when analyzing single-crystal samples. The structural data of AgCoF$_5$ were determined using the powder X-ray diffraction technique on polycrystalline samples. This approach was supported by incorporating computational methods at various stages of the structure determination process. It is therefore essential to bear in mind that the positional data of the light atoms may exhibit subtle degrees of approximation.

**Table SI3.** Structural parameters of AgCoF$_5$ from experimental and various theoretical methods.

| Methods | a [Å] | b [Å] | c [Å] | V [Å$^3$] | β [°] |
|---|---|---|---|---|---|
| Rietveld | 7.274 (2) | 7.628(2) | 7.529(2) | 375.580(19) | 115.976(4) |
| DFT+U (U = 5 eV) | 7.187 | 7.643 | 7.535 | 375.305 | 114.948 |
| DFT+U (U = 8 eV) | 7.157 | 7.607 | 7.508 | 370.475 | 115.003 |
| SCAN | 7.269 | 7.749 | 7.586 | 382.619 | 116.431 |
| HSE06 | 7.201 | 7.661 | 7.527 | 376.855 | 114.819 |

**Table SI4**. Bond lengths and angles in the AgCoF$_5$ structure: Rietveld and theoretical methods. These theoretical calculations are based on idealized conditions (p,T → 0), which leads to some inherent discrepancies compared to experimental data obtained under ambient conditions.

| Methods | dM-M [Å] | dM-F [001] [Å] | dM-F (001) [Å] | Bond angle [°] |
|---|---|---|---|---|
| Rietveld | Co-Ag: 3.637(10), 3.814(10)<br>Ag-Ag and Co-Co: 3.765(10) | Ag-F: 2.562(12)<br>Co-F: 1.905(6) | Ag-F: 2.090(16), 2.052(16)<br>Co-F: 1.827(16), 1.921(13) | Co-F-Co: 162.0(3)° [001];<br>Ag-F-Ag: 107.6(5)° [001]<br>Co-F-Ag: 158.9(12)° [010];<br>130.1(8)° [100]; |
| DFT+U (U = 5 eV) | Co-Ag: 3.593, 3.822<br>Ag-Ag and Co-Co: 3.768 | Ag-F: 2.587<br>Co-F: 1.916 | Ag-F: 2.071, 2.056<br>Co-F: 1.816, 1.943 | Co-F-Co: 158.9° [001];<br>Ag-F-Ag: 107.9° [001]<br>Co-F-Ag: 158.9° [010];<br>127.9° [100]; |
| DFT+U (U = 8 eV) | Co-Ag: 3.579, 3.803<br>Ag-Ag and Co-Co: 3.754 | Ag-F: 2.571<br>Co-F: 1.911 | Ag-F: 2.053, 2.043<br>Co-F: 1.817, 1.943 | Co-F-Co: 158.9° [001];<br>Ag-F-Ag: 108.4° [001]<br>Co-F-Ag: 158.6° [010];<br>127.7° [100]; |
| SCAN | Co-Ag: 3.635, 3.874<br>Ag-Ag and Co-Co: 3.793 | Ag-F: 2.563<br>Co-F: 1.919 | Ag-F: 2.081, 2.085<br>Co-F: 1.826, 1.926 | Co-F-Co: 162.4° [001];<br>Ag-F-Ag: 109.1° [001]<br>Co-F-Ag: 164.1° [010];<br>130.1° [100]; |
| HSE06 | Co-Ag: 3.600, 3.831<br>Ag-Ag and Co-Co: 3.763 | Ag-F: 2.608<br>Co-F: 1.910 | Ag-F: 2.070, 2.059<br>Co-F: 1.822, 1.928 | Co-F-Co: 160.2° [001];<br>Ag-F-Ag: 106.9° [001]<br>Co-F-Ag: 159.6° [010];<br>129.1° [100]; |

## SI.III Comparison of Rietveld fits for *C*2/*c* and *C*2/*m* space groups

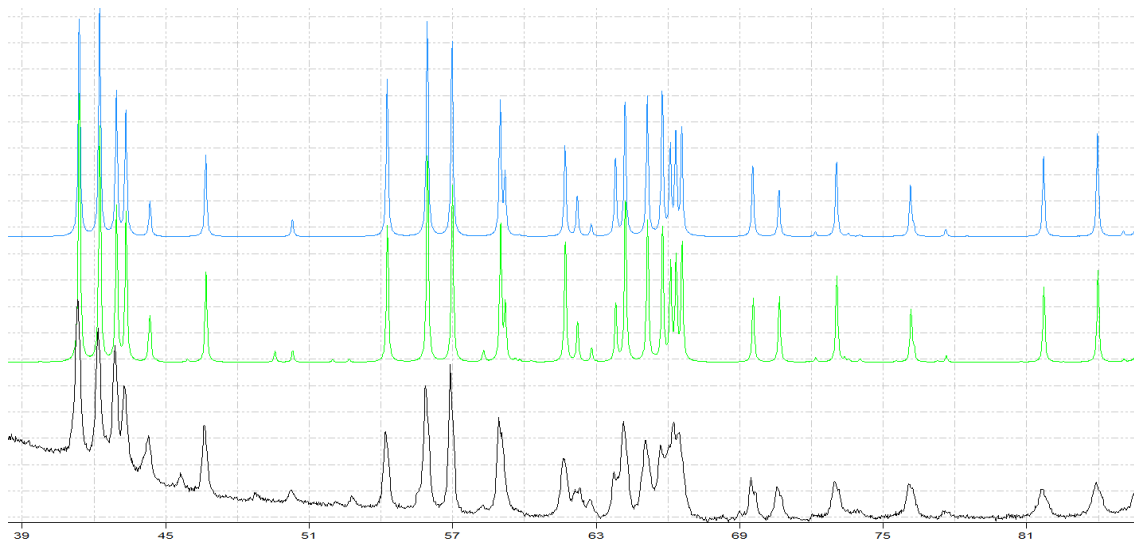

**Figure SI3.** Comparison between experimental diffractogram of **S1** (black line), model diffractogram for AgCoF$_5$ structure in *C*2/*c* (green line) and *C*2/*m* structure (blue line).

**Table SI5.** Structure and fitting parameters of AgCoF$_5$ from the very first cycle of Rietveld refinement process of C2/c and C2/m structures, based on experimental diffractogram from **S1**.

| AgCoF$_5$ group space | Parameters | | | | | Fitting parameters | | |
|---|---|---|---|---|---|---|---|---|
| | a [Å] | b [Å] | c [Å] | V [Å$^3$] | β [°] | GoF | Rp | wRp |
| *C*2/*c* (Z = 4) | 7.276 | 7.629 | 7.531 | 375.803 | 115.981 | 2.56 | 1.63 | 2.59 |
| *C*2/*m* (Z = 2) | 7.276 | 7.629 | 3.765 | 187.886 | 115.981 | 2.91 | 1.88 | 2.95 |

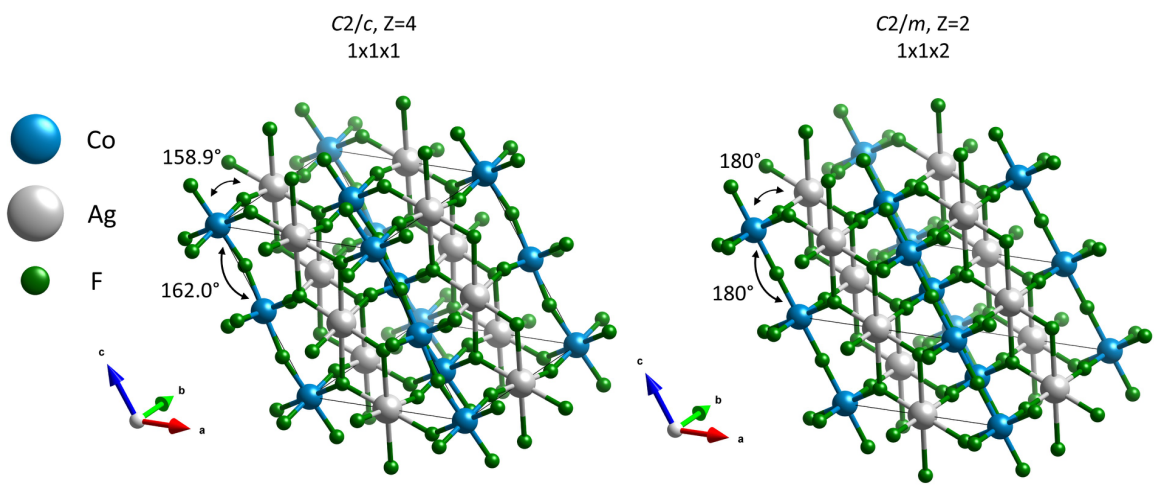

**Figure SI4.** AgCoF$_5$ in C2/c (left) and C2/m (right, 1x1x2 supercell) space group. The unit cell is drawn as a solid line.

# SI.IV Experimental and theoretical phonon frequencies of AgCoF$_5$

**Table SI6.** Phonon vibration frequencies and their symmetries from DFT+U computations and band positions identified in IR and Raman spectra for AgCoF$_5$. Darkened columns indicate bands that are not observable due to symmetry constraints (forbidden by selection rules). Band intensities are categorized as follows: vs - very strong, s - strong, m - medium, w - weak, vw - very weak, sh - shoulder. "Silent" in shaded columns refers to non-active bands. "---" indicates the absence of the corresponding band in the experimental spectra. The indication "n.d." refers to band positions outside the spectroscopes' measurement range. Positions in cm$^{-1}$.

| # | DFT+U | Symmetry | IR | RAMAN | # | DFT+U | Symmetry | IR | RAMAN |
|---|---|---|---|---|---|---|---|---|---|
| 1 | 570 | B$_g$ | | 591vs | 22 | 219 | A$_u$ | | silent |
| 2 | 555 | A$_u$ | silent | | 23 | 215 | B$_u$ | 218m | |
| 3 | 551 | B$_u$ | 548sh | | 24 | 215 | A$_u$ | | silent |
| 4 | 534 | A$_g$ | | --- | 25 | 193 | A$_u$ | | silent |
| 5 | 519 | B$_u$ | 510vs | | 26 | 182 | B$_u$ | 177m | |
| 6 | 471 | B$_g$ | | 491m | 27 | 178 | A$_u$ | | silent |
| 7 | 469 | A$_u$ | silent | | 28 | 176 | B$_g$ | | 173sh |
| 8 | 452 | B$_u$ | 452vs | | 29 | 165 | B$_u$ | --- | |
| 9 | 393 | A$_u$ | silent | | 30 | 156 | A$_u$ | | silent |
| 10 | 391 | A$_g$ | | 393sh | 31 | 139 | B$_u$ | 141vw | |
| 11 | 376 | B$_g$ | | 367sh | 32 | 110 | A$_g$ | | 114vs |
| 12 | 352 | B$_u$ | 356s | | 33 | 102 | B$_g$ | | 100sh |
| 13 | 348 | A$_u$ | silent | | 34 | 98 | A$_u$ | | silent |
| 14 | 314 | A$_u$ | silent | | 35 | 95 | B$_u$ | 92w | |
| 15 | 303 | B$_u$ | --- | | 36 | 79 | A$_g$ | | n.d. |
| 16 | 284 | B$_g$ | | 282sh | 37 | 64 | A$_u$ | | silent |
| 17 | 274 | A$_g$ | | 273sh | 38 | 56 | B$_u$ | n.d. | |
| 18 | 266 | A$_g$ | | 259m | 39 | 45 | B$_g$ | | n.d. |
| 19 | 245 | B$_u$ | 253m | | 40 | -1 | B$_u$ | | |
| 20 | 242 | B$_g$ | | 241sh | 41 | -1 | A$_u$ | | |
| 21 | 220 | A$_g$ | | 223vw | 42 | -2 | B$_u$ | | |

**Table SI7** Combination vibrations and overtones observed in the spectra in Figure 5. Positions in cm$^{-1}$.

| IR [cm$^{-1}$] | RAMAN [cm$^{-1}$] | ASSIGNMENT |
|---|---|---|
| 110 | | 111 B$_u$ -> 45 + 56 (IR$_{Bu}$+R$_{Bg}$) |
| 272 | | 291 (B$_u$) -> 177 + 114 (IR$_{Bu}$ + R$_{Ag}$) |
| 630 | | 624 (B$_u$) -> 510 + 114 (IR$_{Bu}$ + R$_{Ag}$) |
| | 802 | 808 (A$_g$) -> 452 + 356 (IR$_{Bu}$+IR$_{Bu}$) |
| | 1074 | 1082 (A$_g$) -> 591 + 491 (R$_{Bg}$ + R$_{Bg}$) |
| | 1229 | Overtone (2·B$_g$ = 2·591 cm$^{-1}$) |

## SI.V Magnetic data

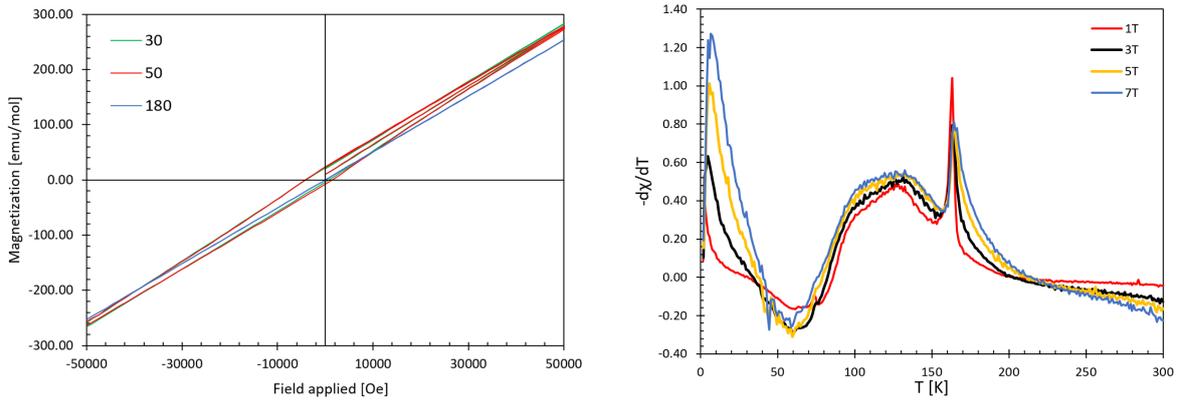

**Figure SI5.** Magnetization curves at 30 K, 50 K, 180 K for **S1** as a function of the applied field (left) and first derivatives of the sample magnetization (-dχ/dT) as a function of T dependence on the left, measured at different fields.

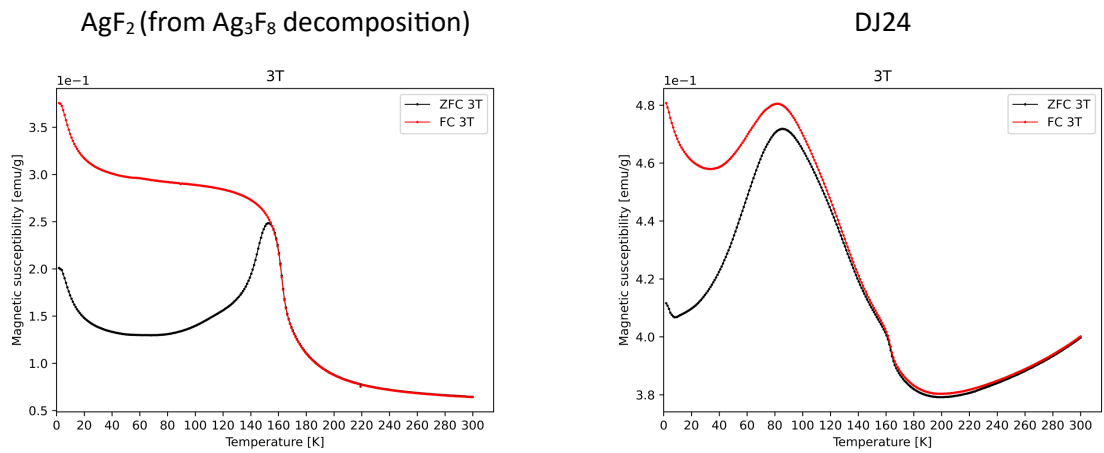

**Figure SI6.** Magnetization for $AgF_2$ (left) and S1 (right) – mainly $AgCoF_5$ with $AgF_2$ traces at 30 kOe vs temperature [K].

A simple subtraction of the contribution of $AgF_2$ to the overall magnetic response of the sample does not eliminate the feature at 163 K (see **Figure SI6**), which is characteristic of the magnetic transition in silver(II) difluoride, from the susceptibility plot. This leads to considerable complications when analyzing the measurement results and makes it impossible to fit a magnetic model. In addition, there are many other complications, including: 1) different magnetic spins for magnetic ions; 2) low-dimensionality of magnetic interactions.

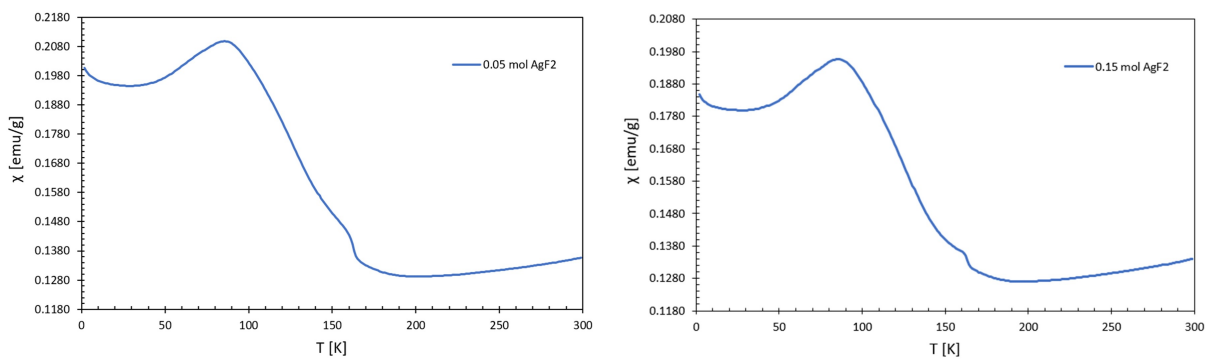

**Figure SI7**. Magnetic susceptibility at 10k Oe of sample 1 (**S1**) after subtracting the contribution of 0.05 mol% (left) and 0.15 mol% (right) of AgF$_2$ to overall magnetic response.

## SI.VI Results of superexchange calculations

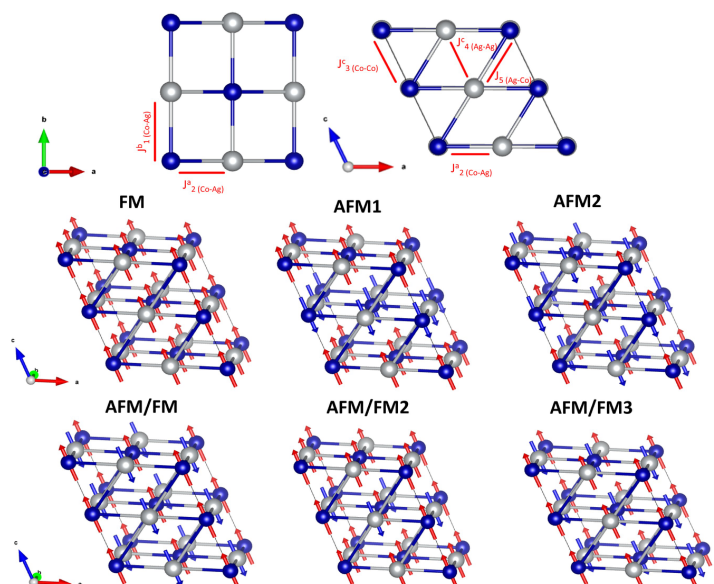

**Figure SI8.** Superexchange paths (top) and six possible spin states for AgCoF$_5$. The fluorine atoms have been omitted for clarity. Blue and red arrows indicate opposite spin directions. AFM2 is the spin state with the lowest energy.

**Table SI8.** Superexchange constants determined with the DFT+U, SCAN and HSE06 methods.

| Method | $J^b_{1(Co-Ag)}$ [010] [meV] | $J^a_{2(Co-Ag)}$ [100] [meV] | $J^c_{3(Co-Co)}$ [001] [meV] | $J^c_{4(Ag-Ag)}$ [001] [meV] | $J_{5(Ag-Co)}$ [101] [meV] | Magnetic moments [$\mu_B$] Ag | Co |
|---|---|---|---|---|---|---|---|
| DFT+U ($U_{Ag}$=5 eV) | -47.74 | -6.53 | -8.28 | -1.11 | -1.23 | +/-3.22 | +/-0.58 |
| DFT+U ($U_{Ag}$=8 eV) | -39.25 | -6.52 | -8.21 | -0.37 | -1.36 | +/-3.24 | +/-0.66 |
| SCAN | -62.02 | -4.77 | -10.28 | -1.34 | -1.91 | +/-3.09 | +/-0.54 |
| HSE06 | -39.49 | -6.74 | -7.96 | 0.70 | -0.52 | +/-3.47 | +/-0.73 |

## SI.VII References

[1] A. Tressaud, J. M. Dance, J. Ravez, J. Portier, P. Hagenmuller, J. B. Goodenough, *Materials Research Bulletin* **1973**, *8*, 1467–1477.
[2] G. Ferey, R. Pape, *Acta Cryst.* **1978**, 1084–1091.
[3] U. Müller, R. Hoppe, *Z. Anorg. Allg. Chem.* **1990**, *583*, 205–208.
[4] J. Graulich, W. Massa, D. Babel, *Z. anorg. allg. Chem.* **2003**, *629*, 365–367.
[5] K. K. Wu, I. D. Brown, *Materials Research Bulletin* **1973**, *8*, 593–598.
[6] S. M. Eicher, J. E. Greedan, *Journal of Solid State Chemistry* **1984**, *52*, 12–21.
[7] R. Von Der Mühll, S. Andersson, J. Galy, *Acta Crystallogr B Struct Crystallogr Cryst Chem* **1971**, *27*, 2345–2353.
[8] R. Georges, J. Ravez, R. Olazcuaga, P. Hagenmuller, *Journal of Solid State Chemistry* **1974**, *9*, 1–5.
[9] R. von der Mühll, F. Daut, J. Ravez, *Journal of Solid State Chemistry France* **1973**, *8*, 206–212.
[10] A. Tressaud, J. M. Parenteau, J. M. Dance, J. Portier, P. Hagenmuller, **1973**, *8*.
[11] M. Body, G. Silly, C. Legein, J.-Y. Buzaré, F. Calvayrac, P. Blaha, *Journal of Solid State Chemistry* **2005**, *178*, 3655–3661.
[12] J. Ravez, M. Darriet, R. Von der Mühll, P. Hagenmuller, *Journal of Solid State Chemistry* **1971**, *3*, 234–237.
[13] J. Grannec, J. Ravez, *C. R. Acad. Sc. Série C* **1970**, *270*, 2059–2061.